\documentstyle[prl,aps]{revtex}

\def\la{\langle}
\def\ra{\rangle}

\def\lb{\lbrack}
\def\rb{\rbrack}

 \def\Slash#1{
  \begin{picture}(5,6)(0,0)
  \put(-.7,-1.2){\line(5,6)6}
  \end{picture}
  \kern-.8em#1}
 \def\slash#1{
  \begin{picture}(5,6)(0,0)
  \put(-1.5,-1.7){\line(5,6)5}
  \end{picture}
  \kern-.8em#1}

\def\Tr{\mbox{Tr}}

\def\e{\epsilon}
\def\gg5{\gamma_5}
\def\hg5{\hat{\gamma}_5}
\def\g4{\gamma_4}

\def\U{{\cal U}}

\def\D{{\cal D}}

\def\Qlatmr1{Q_{lat}^{(m=r=1)}}

\def\be{\begin{eqnarray}}
\def\ee{\end{eqnarray}}

\def\T{{\cal T}}

\def\bx{{\bf x}}
\def\bp{{\bf p}}
\def\t{\tau}

\def\wD{\widetilde{D}}
\def\wPsi{\widetilde{\Psi}}
\def\bmu{\mu^*}

\begin{document}

\draft
 
\title{A dimensionally reduced expression for the QCD fermion determinant at finite
temperature and chemical potential}

\author{David H. Adams}

\address{Instituut-Lorentz for Theoretical Physics, Leiden University, 
Niels Bohrweg 2, NL-2333 CA Leiden, The Netherlands. Email: adams@lorentz.leidenuniv.nl}

\date{March.'04}

\maketitle

\begin{abstract}

A dimensionally reduced expression for the QCD fermion determinant at finite temperature 
and chemical potential is derived which sheds light on the determinant's
dependence on these quantities.
This is done via a partial zeta-regularisation, formally applying a
general formula for the zeta-determinant of a differential operator in one variable
with operator-valued coefficients. 
The resulting expression generalises the known one for the free fermion determinant, 
obtained via Matsubara frequency summation, to the case of general background gauge field; 
moreover there is no undetermined overall factor. Rigorous versions of the result are 
obtained in a continuous time---lattice space setting.
The determinant expression reduces to a remarkably simple form in the low temperature
limit. A program for how to use this to obtain insight into the QCD phase transition
at zero temperature and nonzero density is outlined.

\end{abstract}


\widetext

QCD at finite temperature and density exhibits interesting phase structure; in particular,
at sufficiently high temperature and/or density there is a transition from the usual 
confined, chiral symmetry-broken hadronic phase to a deconfined,
chirally symmetric phase where the quarks and gluons are ``liberated'' and form a
quark-gluon plasma (QGP). The Universe was very likely a QGP for a brief moment after the
Big Bang, and such a phase may also exist at present in the cores of very dense
stars (neutron/quark stars). Furthermore, it is a major aim of current heavy ion 
collision experiments at RHIC and CERN to create a QGP for brief instants
through the high-energy collisions of heavy nuclei. QCD at finite temperature and density
is therefore an exciting field of major current research interest \cite{Alford}.
The equilibrium properties are governed by the QCD grand-canonical partition function
(referred to hereafter simply as the 'partition function'). A crucial quantity
contained in this is the fermion determinant, which encodes the dynamical fermion effects. 
The central problem
in QCD thermodynamics is to determine the dependence of the partition function on the
temperature $T$ and (quark) chemical potential $\mu\,$, and to this end it is highly 
desirable to get as much information as possible on how the fermion determinant depends on 
these quantities. In this paper we derive a ``dimensionally reduced'' expression for the
fermion determinant which sheds light on these dependencies.

The importance of the fermion determinant in this context has been demonstrated in
Lattice QCD studies at low temperature and nonzero chemical potential.
It was found that when the fermion determinant is discarded (i.e in the ``quenched 
approximation'') the onset of the chiral phase transition in the quark number density
and chiral condensate occurs at 
$\mu_o=m_{\pi}/2$ (where $m_{\pi}$ is the lightest meson mass, i.e. the pion mass) 
rather than at the expected critical value $\mu_c=m_N/3$ (where $m_N$ is the lightest
baryon mass, i.e. the nucleon mass) ---see \cite{Kogut97review} and the references
therein. The necessity of including the fermion determinant has also been demonstrated
in another approach based on Random Matrix Theory \cite{Stephanov}. 
The phase transition at zero temperature and nonzero density is of central interest
since, e.g., it is expected to occur in the formation of sufficiently dense stars.
Unfortunately, this region of the phase diagram is not currently accessible by the currently
developed (non-quenched) lattice methods or perturbative methods \cite{Alford}. 
However, there is an indication
that it may be possible to obtain insight into this region by nonperturbative analytic
techniques. Recently, T.~Cohen has derived an interesting expression for the ratio
$det D_A(\mu)/det D_A(0)$ where $det D_A(\mu)$ is the finite temperature fermion determinant 
in a background gauge field $A$ and at chemical potential $\mu$ \cite{Cohen}. In the low
temperature limit it reduces to a remarkably simple form and the fermion
determinant is seen to undergo a transition from $\mu$-independence to $\mu$-dependence
at precisely the critical value $\mu_o$ mentioned above \cite{Cohen}. The zero-temperature 
QCD partition function itself should undergo a transition at the above-mentioned 
critical value $\mu_c$. Its $\mu$-independence in the region $\mu_o<\mu<\mu_c$ must 
therefore involve a subtle cancellation in the functional integral of 
$e^{-S_{YM}(A)}\,\prod_{j=1}^{N_f}det D_A(\mu\,,m_j)$ over gauge fields 
(where $S_{YM}(A)$ is the Yang-Mills action, $N_f$ the number of quark flavours and 
$m_j$ their masses). 
This is intimately connected with the nature of the strong interactions, and
in particular, confinement, in usual QCD (i.e. at $T\!=\!\mu\!=\!0$), since in the absence
of these interactions we would simply have $m_{\pi}/2=m_N/3\,$=lightest quark mass,
i.e. $\mu_o=\mu_c$. 
Therefore, understanding how this 
cancellation works is not only key to understanding the hadronic-to-QGP phase transition
in QCD at zero-temperature and nonzero density but should also give insight into
confinement in usual QCD. 

Understanding this cancellation is therefore 
an interesting and important problem (dubbed the 'baryon Silver Blaze' problem
in \cite{Cohen}). To make progress on this, a necessary first step is to supplement
the determinant ratio expression of \cite{Cohen} with an expression for $det D_A(\mu)$
itself, since we need to know its full gauge field-dependence in order to study the 
aforementioned cancellation in the functional integral over gauge fields. 
(A knowledge of $det D_A(\mu)/det D_A(0)$ does not suffice for this since we cannot
a priori rule out the possiblity that there are important gauge field-dependent factors 
which drop out in the determinant ratio.)
This provides a concrete motivation for the considerations in the present paper. 
The ``dimensionally reduced'' expression for $det D_A(\mu)$ that we obtain, (\ref{10})
below, also reduces to a remarkably simple form in the low $T$ limit 
((\ref{16a})--(\ref{16b}) below) and we identify a gauge field-dependent factor which
drops out in the determinant ratio.
Our results are derived both at the formal continuum level and 
rigorously in a continuous time---lattice space setting. (In the latter setting the evaluation
requires certain choices to be made, resulting in different overall factors in
the determinant expression, and it is a non-trivial question whether these are 
physically equivalent. This issue is intimately connected with a ``universality anomaly''
in Lattice QCD uncovered recently in \cite{simplified}.) The natural next step in this
program, which is currently under investigation, is to consider the strong coupling
limit of the functional integral expression for the QCD partition function, in the
low temperature limit with the fermion determinant expression obtained here, and compare
it with the expression obtained in the strong coupling lattice Hamiltonian framework at
finite chemical potential. (Note that this comparison makes sense in the 
continuous time---lattice space setting.) 
The latter has been studied, e.g., in \cite{Luo} and the 
phase transition at the appropriate $\mu_c$ has been explicitly demonstrated in that 
framework. Using the equivalence between the functional integral and Hamiltonian frameworks 
it should be possible to see how the aforementioned cancellation
in the functional integral over gauge fields for $\mu_o<\mu<\mu_c$ comes about in
the strong coupling limit. The final, and most challenging, step in the program will
then be to extend the understanding of this cancellation to the case of general
couplings. 

The ``dimensional reduction'' method introduced here is also potentially useful
in other contexts and we briefly discuss several of these at the end of the paper.

The QCD partition function with $N_f$ quark flavours with masses $m_j$ and quark chemical
potential $\mu$ can be expressed as a functional integral,
\be
Z(\beta,\mu)=
\int\D A\;e^{-S_{YM}(A)}\,\prod_{j=1}^{N_f}\,\int\D\bar{\psi}_j\D\psi_j
\;e^{-\int_0^{\beta}d\tau\,\int_{space}d^3\bx\,\bar{\psi}_j\lb\gamma_{\nu}
(\partial_{\nu}+A_{\nu})+m_j-\mu\gamma_4\rb\psi_j}
\label{2}
\ee
where $\beta=1/T$. The fermion fields $\psi_j\,$, $\bar{\psi}_j$ (resp. gauge fields
$A_{\nu}$) are required to satisfy anti-periodic (resp. periodic) boundary conditions
in the Euclidean ``time'' variable $\tau\in[0,\beta]$. 
3-space is taken to be a box (whose 
volume $V$ is taken to infinity in the final step of calculating physical quantities).
The spacial b.c.'s will not play a role in our considerations and we leave them unspecified.
The integrals over the fermion fields in (\ref{2}) give the fermion determinants, i.e.
for $m_j\!=\!m$ the integral gives $det D_A(\mu)$ where
\be
D_A(\mu)=\gamma_{\nu}(\partial_{\nu}+A_{\nu})+m-\mu\gamma_4
\label{3}
\ee

In the free field ($A=0$) case, where $det D_0(\mu)$ is the partition function of a 
free Dirac fermion gas, the dependence of the determinant on $\beta$ and $\mu$ is 
well-known: An application of the standard Matsubara frequency summation method gives
\cite{Negele-Kapusta}
\be
det D_0(\mu)=C_0\,\prod_{\bp}\,\lb\,e^{\beta E}\,(1+e^{-\beta(E-\mu)})
(1+e^{-\beta(E+\mu)})\rb^2
\label{4}
\ee
with $\pm E=\pm\sqrt{\bp^2+m^2}$ being the (2-fold degenerate) energy eigenvalues of
$H_0=\gamma_4(\gamma_k\partial_k+m)$ (summation over $k\!=\!1,2,3$ is implied).
From this, expressions for physical quantities such as the energy density 
$\la E\ra/V=-\frac{1}{V}\frac{\partial}{\partial\beta}\,log\,det D_0\,|_{\beta\mu=const}$
and particle number density 
$\la Q\ra/V=\frac{1}{V}\frac{\;\partial}{\beta\partial\mu}\,log\,det D_0$ 
are obtained; in particular one finds for massless fermion the 
well-known results $\la E\ra/V\sim\mu^4$ and $\la Q\ra/V\sim\mu^3$ in the 
large $\beta$ limit. The Matsubara summation produces an undetermined overall factor
$C_0$ in (\ref{4}); this is inconsequential though since it does not involve 
$\beta\,$, $\mu$ or the energies $\pm E$.

We now consider the problem of generalising (\ref{4}) to the case of arbitrary background 
gauge field. The Matsubara summation method can be used to obtain an expression for
the ratio $det D_A(\mu)/det D_A(0)$ \cite{Cohen}; however, it is of limited use for 
$det D_A(\mu)$ itself since, e.g., it produces an undetermined overall factor and we cannot 
exclude a priori the possibility that this factor may depend on the gauge field. 
Therefore we take a different approach. Regarding the spinor fields 
$\psi(\bx,\tau)$ as functions $\Psi(\tau)$ taking values in the vectorspace $W=\{\psi(\bx)\}$
of spinor fields living only in the spacial volume (i.e. the usual Hilbert space of quantum
mechanical wave functions), we re-express $D_A(\mu)$ as
\be
D_A(\mu)=\gamma_4\,(\,{\textstyle \frac{\partial}{\partial\tau}}+H_A(\tau)-\mu)
\label{5}
\ee
where $H_A(\t):W\to W$ is the linear map defined by
\be
H_A(\tau)\psi(\bx)=\lb\gamma_4(\gamma_k(\partial_k+A_k(\bx,\tau))+m)+A_4(\bx,\tau)\rb\psi(\bx)
\label{6}
\ee
In this way $D_A(\mu)$ can be viewed as a 
differential operator in one variable $\tau\in[0,\beta]$
acting on $W$--valued anti-periodic functions. An expression for its determinant can then
be obtained by application of a zeta-regularised determinant formula in \cite{BFK} 
(see also \cite{Forman} for related and overlapping results). Recall that the zeta-determinant
\cite{RS} of an operator $D$ is defined by $det_{\zeta}D=\mbox{''$\prod_j\lambda_j$''}
\equiv e^{-\zeta_D'(0)}$ where $\{\lambda_j\}$ are the eigenvalues of $D$ and $\zeta_D(s)$
is the zeta function defined by $\zeta_D(s)=\sum_j\frac{1}{\lambda_j^s}=Tr(D^{-s})$. 
Under certain ellipticity conditions $\zeta_D(s)$ is a well-defined
smooth function of the complex parameter $s$ when $Re(s)$ is sufficiently large, and can be
analytically continued to a meromorphic function in the whole complex plane which is regular 
at $s=0$, so that $\zeta_D'(0)$ and hence $det_{\zeta}D$ are well-defined. The zeta-determinant
formula of \cite{BFK}, in its most straightforward form, is for elliptic differential operators 
in one variable $t\in[0,1]$ acting on periodic functions taking values in some vectorspace.   
To apply it in the present case, we note that the spectrum of $D_A(\mu)$ is the same as
that of the operator $\wD_A(\mu)=\gamma_4\,(\,\frac{1}{\beta}\frac{\partial}{\partial t}
+H_A(\beta t)-\mu-\frac{i\pi}{\beta})$ acting on $W$--valued {\em periodic} functions
$\wPsi(t)\,$, $t\in[0,1]\,$: It is easy to check that $\Psi(\tau)$ is an eigenfunction
for $D_A(\mu)$ with eigenvalue $\lambda$ if and only if 
$\wPsi(t)=e^{i\pi t}\,\Psi(\beta t)$ is an eigenfunction for $\wD_A(\mu)$ with the same
eigenvalue. Hence $det D_A(\mu)=det\wD_A(\mu)$, and the result of \cite{BFK}
can be applied to evaluate the latter determinant. 
Writing $\wD_A(\mu)=L_1(t)\,\frac{\;d}{idt}+L_0(t)$ where 
$L_1(t)=\frac{i}{\beta}\gamma_4\;$, 
$\,L_0(t)=\gamma_4\,(\,H_A(\beta t)-\mu-\frac{i\pi}{\beta})\,$, 
and formally setting $N=dim W$, Theorem 1 of \cite{BFK} 
gives
\be 
det D_A(\mu)=det\wD_A(\mu)=(-1)^N\,S_{\theta}(L_1,L_0)\,R(L_1,L_0)\,
det({\bf 1}-\U(L_1,L_0))
\label{7}
\ee
where the ingredients are as follows. Consider the equation $\wD_A(\mu)\wPsi(t)=0$ without
boundary conditions on $\wPsi(t)$. The solutions are determined from their initial values
via an evolution operator: $\wPsi(t)=\widetilde{\U}(t)\wPsi(0)$. Then 
$\U(L_1,L_0):=\widetilde{\U}(1)=T\,e^{-i\int_0^1L_1(t)^{-1}L_0(t)\,dt}$ ($T$=$t$-ordering).
The remaining ingredients are given by 
$R(L_1,L_0):=e^{\frac{i}{2}\int_0^1Tr(L_1(t)^{-1}L_0(t))\,dt}$ and $S_{\theta}(L_1,L_0):=
(det\,\Gamma_{\theta})\,e^{\frac{i}{2}\int_0^1Tr(\Gamma_{\theta}(t)L_1(t)^{-1}L_0(t))\,dt}$
with $\Gamma_{\theta}(t)$ as defined in the following. Note that for $\lambda\in{\bf C}$ the
definition of $\lambda^{-s}$ depends on a choice of cut in the complex plane; the same is
true for $\zeta_D(s)$, and hence the zeta-determinant $det_{\zeta}D$ also depends on such a 
choice. In the formula (\ref{7}) this choice is represented by $\theta$. Specifically,
choose $\theta\in{\bf R}$ so that the line $\{re^{i\theta}\,|\,r\in{\bf R}\}$ does not 
contain any eigenvalue of $L_1(t)$. Then the argument $\alpha$ of an eigenvalue of 
$L_1(t)$ satisfies either $\theta<\alpha<\theta+\pi$ or $\theta+\pi<\alpha<\theta+2\pi$.
Let $\Pi_{\theta}^+(t)$ and $\Pi_{\theta}^-(t)$ denote the projections onto the subspaces of
$W$ spanned by the eigenvectors of $L_1(t)$ for which the arguments of the corresponding
eigenvalues are in the intervals $\rb\theta,\theta+\pi\lb$ and $\rb\theta+\pi,\theta+2\pi\lb$,
respectively. Then $\Gamma_{\theta}(t):=\Pi_{\theta}^+(t)-\Pi_{\theta}^-(t)$.
Note that the matrix for $\Gamma_{\theta}(t)$ in the basis of eigenvectors for $L_1(t)$ is
diagonal with entries $\pm1$, so $det\,\Gamma_{\theta}(t)=\pm1$ independent of $t$. In the
present case $L_1(t)=\frac{i}{\beta}\gamma_4$ and it is easy to see that 
$\Gamma_{\theta}=\pm\gamma_4$ with sign ``$+$'' for $-\pi/2<\theta<\pi/2$ and 
``$-$'' for $\pi/2<\theta<3\pi/2$. 
Now, inserting the above expressions for $L_1$ and $L_0(t)$ into $\U(L_1,L_0)\,$,
$S_{\theta}(L_1,L_0)\,$, and $R(L_1,L_0)\,$, and inserting these into (\ref{7}), and noting
the following easy consequences of the definition (\ref{6}) of $H_A(\tau)$ (which follow
from the vanishing of the traces of $\gamma_{\nu}\,$, $\,\gamma_4\gamma_{\nu}$ and the 
anti-hermitian matrix $A_4(\bx,\t)$):
\be
Tr(H_A(\tau)-\mu)&=&-\mu N \label{8} \\
Tr(\Gamma_{\theta}(H_A(\tau)-\mu))&=&Tr(\pm\gamma_4(H_A(\tau)-\mu))=\pm mN \label{9}
\ee
we arrive after a little calculation at the result
\be
det D_A(\mu)=C\,e^{-\beta\mu N/2}\,det({\bf 1}+e^{\beta\mu}\,\U_A(\beta))
\label{10}
\ee
where
\be 
\U_A(\beta)=\T\,e^{-\int_0^{\beta}H_A(\tau)\,d\tau}
\label{11}
\ee
and $C=e^{\pm\beta mN/2}$.
($\T$=$\tau$-ordering and we have used the fact that $N=dim W$ is formally even,
since spinor fields have an even number of components, hence $(-1)^N=1$.)
We will discuss below how variants of this approach can produce different expressions
for the overall factor $C$ in (\ref{10}). In the present case $C=e^{\pm\beta mN/2}$ is gauge 
field-independent -- its only effect is to give an overall shift in quantities obtained
by taking derivatives of $logZ$ with respect to $\beta$ or $m$ --
so the the factors $C\,e^{-\beta\mu N/2}$ in (\ref{10})
are physically inconsequential and can be absorbed into a normalisation of the 
QCD partition function; hence the physics of the fermion determinant is described solely by 
$det({\bf 1}+e^{\beta\mu}\,\U_A(\beta))$. Thus (\ref{10}) is a {\em dimensionally reduced} 
expression for the fermion determinant, showing that the fermion action in (\ref{2}) can be 
replaced by a non-local action for fermion fields living in one spacetime dimension lower 
(i.e. living only in the spacial volume). Moreover, it provides the promised generalisation 
of (\ref{4}) to the case of arbitrary background gauge field:
In the free field case, $\U_0(\beta)=e^{-\beta H_0}$ and it is easy to check that (\ref{10}) 
reproduces (\ref{4}) with $C_0=C$. 

The result (\ref{10}) also provides another way to relate the functonal integral and 
Hamiltonian frameworks for quantum field theories with fermions, alternative to the usual
relation based on transfer matrices which requires a time discretisation \cite{Creutz-transfer}. 
In the free field case this is easily
seen using a standard algebraic rewrite of the determinant in the right-hand side
of (\ref{10}): Defining the extension of an operator $Q$ on $W$ to an operator $Q^{\wedge}$ 
on the Clifford algebra $\oplus_{p=0}^NW^{{\wedge}p}$ ($\simeq$fermionic Fock space)
by $Q^{\wedge}(\psi_1\wedge\cdots\wedge\psi_p)=(Q\psi_1)\wedge\cdots\wedge(Q\psi_p)$
we have $det({\bf 1}+e^{-\beta(H_0-\mu)})=Tr_{Fock}((e^{-\beta(H_0-\mu)})^{\wedge})$
and the latter can be identified with $Tr(e^{-\beta({{\cal H}}-\mu{{\cal N}})})$
where ${\cal H}$ and ${\cal N}$ are the fermionic Hamiltonian and number operators,
respectively, on the Fock space. (In this description the vacumm state has
{\em unfilled} Dirac sea, i.e. it consists of no particles of either positive or negative
energy. Vacuum subtractions then need to be done ``by hand'' to obtain
physical results.)
Relating the functional integral and Hamiltonian approaches via (\ref{10}) in the
general case where gauge fields are present is more involved
and will be discussed in a seperate paper.

The above derivation of (\ref{10}) is formal since $W=\{\psi(\bx)\}$ is infinite-dimensional. 
However, a rigorous version can be obtained by putting the spacial volume on a lattice, 
i.e. by working in a continuous time---lattice space setting where $W$ is the 
finite-dimensional vectorspace of spinor fields living on the sites of the spacial lattice.
The Dirac operator in that setting was discussed in \cite{simplified} (see eq.(7) of that 
paper); it is obtained from the above expression (\ref{5})--(\ref{6}) for $D_A(\mu)$
by replacing $\partial_k+A_k\to\nabla_k^A$ (=the covariant finite difference operator
constructed with the link variables of the gauge field on the spacial lattice)
and $m\to M_A(\tau)=m+\frac{r'}{2a'}\Delta_{space}^A(\tau)\,$, where 
$\frac{r'}{2a'}\Delta_{space}^A(\tau)$ is the spacial Wilson term as defined in 
\cite{simplified}. (We continue to denote the gauge field by $A$ in the 
continuous time---lattice space setting even though its spacial components are link 
variables.)
The derivation of the dimensionally reduced expression for the fermion
determinant then goes through as above but with one small change: $m\to M_A(\tau)$ in the 
right-hand side of (\ref{9}), and consequently $C=e^{\pm\beta mN/2}$ gets replaced 
in (\ref{10}) by
\be
C_{\pm}(A)=e^{\pm\frac{1}{2}\int_0^{\beta}Tr\,M_A(\tau)\,d\tau}
\label{x1}
\ee
This modification is significant: $C_{\pm}(A)$ is gauge field-dependent and 
the indeterminacy of the sign in the argument of the exponential 
therefore constitutes an inconsistency
in continuous time---lattice space QCD when the fermion determinant is defined via
zeta-regularisation. There is a way to avoid this inconsistency though. 
Since $\bar{\psi}=\psi^*\gamma_4$ the fermion determinant can be expressed as
$det(\gamma_4D_A(\mu))$ (in fact this is the starting point for the usual evaluation of 
the fermion determinant via Matsubara frequency summation in the free field case).
Although formally $det(\gamma_4D_A(\mu))$ coincides with $det D_A(\mu)$, the expressions
obtained from zeta-regularisation turn out to be different. The difference originates from
a difference in $\Gamma_{\theta}\,$: For $\gamma_4D_A(\mu)$ we find
$\Gamma_{\theta}=\pm1$ (with the sign depending on 
$\theta$ in the same way as previously), and consequently eq.(\ref{9}) becomes
$Tr(\Gamma_{\theta}(H_A(\tau)-\mu))=Tr(\pm(H_A(\tau)-\mu))=\mp\mu N$. The result is that
$det(\gamma_4D_A(\mu))$ is given by (\ref{10}) with $C=e^{\mp\beta\mu N/2}\,$, both
in the formal continuum setting and in the regularised (spacial lattice) setting.
Since this $C$ is gauge field-independent the sign indeterminacy does not matter and
the factor can be absorbed into a normalisation of the QCD partition function.

Thus in the continuous time---lattice space setting the overall factor
$C$ in the dimensionally reduced expression (\ref{10}) for the fermion determinant
depends on choices made in the evaluation: it can be either
$C_+(A)\,$, $C_-(A)$ or $e^{\mp\beta\mu N/2}$. Clearly, the question of whether these
choices are physically equivalent corresponds to the question of whether or not
the gauge field-dependent factor $C_{\pm}(A)$ is physically consequential.
This same question has recently arisen in an investigation of universality 
in Lattice QCD \cite{simplified}. In fact (\ref{10}) has a very similar structure 
to a Lattice QCD fermion determinant expression obtained previously by an algebraic method 
in \cite{Gibbs} (see also \cite{others}). Its continuous time limit was evaluated 
in \cite{simplified} for various versions of the Wilson/naive lattice fermion formulations,
and, depending on the choice of lattice formulation, the limit was found to coincide with 
(\ref{10}) with $C=C_+(A)$ or $C=1$ (modulo some physically inconsequential factors).
The factor $C_+(A)$ was thus seen to represent a ``universality anomaly'' in that case.

How serious a problem is this in the present case?
Zeta-regularisation has become a rather standard
technique which seems to have always given sensible, consistent results in the cases where 
it has been applied (see, e.g., \cite{Elizalde-book}), while lattice regularisation is a 
well-established and successful approach at least in QCD. It would therefore be an unpleasant 
shock to find that the combination of these regularisations in the present setting leads to 
a physically ambiguous or inconsistent result for the fermion determinant. 
To determine whether this is actually the case one needs to determine the physical 
significance, or lack thereof, of the troublesome factor $C_{\pm}(A)$. To this end,  
recall that $M_A(\tau)=m+\frac{r'}{2a'}\Delta_{space}^A(\tau)$ in (\ref{x1}).
One could argue that this factor is physically inconsequential when one goes on to take the
continuous space limit ($a'\to0$) since the spacial Wilson term 
$\frac{r'}{2a'}\Delta_{space}^A$ formally vanishes in this limit. This is a delicate issue 
though, since $Tr\frac{1}{a'}\Delta_{space}$ actually diverges in this limit (the largest
eigenvalue is $\sim\frac{1}{a'}$). It is tempting to interpret this divergence as being due
to the spacial ``fermion doubler'' modes, which get masses $\sim\frac{1}{a'}$ from the 
spacial Wilson term and are supposed to decouple in the spacial continuum limit. 
But the situation may be more complicated than this and further study is required 
to clarify this issue.

In any case, the dimensionally reduced expression (\ref{10}) sheds light on the $\beta$-
and $\mu$-dependence of the fermion determinant and we study this in the remainder of the
paper, focusing in particular on the $\mu$-dependence. 
Further details of our calculations, along with some additional results, are given 
in \cite{prep}. We begin by noting some general features.
The operator $\U_A(\beta)$ given in (\ref{11}) has the properties
\be
(\U(\beta)^*)^{-1}&=&(\gamma_4\gamma_5)^{-1}\,\U(\beta)\,(\gamma_4\gamma_5)
\label{13} \\
det\,\U_A(\beta)&=&1 \label{14}
\ee
The derivation of the first of these is postponed to \cite{prep}; it immediately
gives $|det\,\U_A(\beta)|^2\!=\!1$, but to derive (\ref{14}) we need to show the absence 
of a phase factor and this requires a direct calculation: $\frac{d}{d\beta}\,det\,\U(\beta)
=det\,\U(\beta)\,Tr(\U(\beta)^{-1}\frac{d}{d\beta}\,\U(\beta))
=det\,\U(\beta)\,\Tr(\U(\beta)^{-1}(-H(\beta))\,\U(\beta))
=-det\,\U(\beta)\,Tr\,H(\beta)\,$; this vanishes by (\ref{8}), hence
$det\,\U(\beta)=det\,\U(0)=1$ as claimed. Using (\ref{13})--(\ref{14}) it is straightforward 
to see that the dimensionally reduced expression (\ref{10}) satisfies the standard relation
\be
det D_A(\mu)^*=det D_A(-\bmu)\,,
\label{12}
\ee
(which follows at the formal level from $D_A(\mu)^*=\gamma_5\,D_A(-\bmu)\,\gamma_5$), 
showing that $det D_A(\mu)$ is complex-valued in general but real for purely imaginary $\mu$.

Specialising to real $\mu$ we consider now the large $\mu$ limit of the fermion determinant. 
In this limit the fermion lagrangian in (\ref{2}) is dominated by the term 
$\mu\bar{\psi}\gamma_4\psi\,$; thus, formally, the dynamical fermion effects disappear from 
the QCD partition function in this limit and the thermodynamics is that of the pure gauge 
theory. This should correspond to the fermion determinant becoming independent of the gauge
field in the large $\mu$ limit, and the question of whether this happens
can be investigated directly from the
dimensionally reduced expression derived in this paper: As a consequence of
(\ref{14}) we have $det({\bf 1}+e^{\beta\mu}\,\U_A(\beta))\approx e^{\beta\mu N}$ in the 
large $\mu$ limit, and it follows that in the formal continuum setting the expression 
(\ref{10}) does indeed become independent of the gauge field. However, in the regularised
(spacial lattice) setting a residual gauge field dependence remains, contained in the 
notorious factor $C_{\pm}(A)$. Thus the intuitive expectation of
gauge field independence in the large $\mu$ limit is not entirely realised in this case.

To study the $\mu$-dependence of the fermion determinant in more detail it is useful to
re-express (\ref{10}) in terms of the eigenvalues of
$\gamma_4D_A(0)=\frac{\partial}{\partial\t}+H_A(\t)$ as follows. 
Consider the eigenvalue equation
\be
\frac{\partial\Psi}{\partial\t}+H_A(\t)\,\Psi(\t)&=&(\lambda+\frac{i\pi}{\beta})\Psi(\t)
\quad,\qquad\ \Psi(\beta)=-\Psi(0)
\label{x8}
\ee
A little analysis shows that the solutions are given by
$\Psi(\t)=e^{(\lambda+\frac{i\pi}{\beta})\t}\,\U_A(\t)\,\Psi(0)$ with
$\U_A(\beta)\,\Psi(0)=e^{-\beta\lambda}\,\Psi(0)$. From this we see that the eigenvalues 
of $\frac{\partial}{\partial\t}+H_A(\t)$ come in equivalence classes
$\{\{\lambda+\frac{i\pi}{\beta}(2n+1)\}_{n\in{\bf Z}}\}$ which are in one-to-one
correspondence with the eigenvalues $\{e^{-\beta\lambda}\}$ of $\U_A(\beta)$. 
The representative $\lambda$ for the equivalence class
$\{\lambda+\frac{i\pi}{\beta}(2n+1)\}_{n\in{\bf Z}}$ can be fixed by the condition 
$Im(\lambda)\in\lb-\pi/\beta,\pi/\beta\lb$. A known property of the eigenvalues that
we exploit in the following is that the $\lambda$'s with $Re(\lambda)\ne0$ come in pairs 
$(\lambda,-\lambda^*)$ (a derivation of this is provided in \cite{prep}).
Using this in (\ref{10}) we obtain
\be
det D_A(\mu)&=&C\,\prod_{\lambda}
e^{-\beta\mu/2}\,(1+e^{\beta\mu}\,e^{-\beta\lambda}) \nonumber \\
&=&C\,\Big\lb\,\prod_{Re(\lambda)=0}
e^{-\beta\mu/2}\,(1+e^{\beta(\mu-\lambda)})\Big\rb\Big\lb\,\prod_{Re(\lambda)>0}
e^{-\beta\mu}\,(1+e^{\beta(\mu-\lambda)})(1+e^{\beta(\mu+\lambda^*)})\Big\rb
\label{15}
\ee
Note that the last square-bracketed factor can be rewritten as 
$\prod_{Re(\lambda)>0}e^{\beta\lambda^*}(1+e^{-\beta(\lambda-\mu)})
(1+e^{-\beta(\lambda^*+\mu)})$, which has the same structure as the earlier free field
expression (\ref{4}). This can now be used to obtain an expression for the fermion determinant 
in the limit of low temperature (i.e. large $\beta$). After some manipulations we find 
\be
det D_A(\mu)&\stackrel{\beta\to\infty}{\approx}&C\,e^{-\beta E_{sea}(A)}\,\epsilon(A)\,
\prod_{|Re(\lambda)|<|\mu|}e^{\beta(|\mu|-|Re(\lambda)|)/2}\,
e^{-i\beta\frac{\mu}{|\mu|}\,Im(\lambda)/2}
\qquad\ (\mu\ne0) \label{16a} \\
det D_A(0)&\stackrel{\beta\to\infty}{\approx}&C\,e^{-\beta E_{sea}(A)}\,\epsilon(A)\,
\prod_{Re(\lambda)=0}2\,cos(\beta\,Im(\lambda)/2) \label{16b}
\ee
where $E_{sea}(A):=\sum_{Re(\lambda)<0}Re(\lambda)$ is the sum of the negative quasi-energies
(this is the natural generalisation of the sum of energies of the negative energy states in 
the Dirac sea arising in the free fermion case), and
\be
\epsilon(A):=\prod_{\lambda}\,e^{-i\beta\,Im(\lambda)/2}\,.
\label{17}
\ee
The latter is a gauge field-dependent sign factor: 
$\epsilon(A)^2=\prod_{\lambda}\,e^{-i\beta\,Im(\lambda)}=\prod_{\lambda}e^{-\beta\lambda}
=det\,\U_A(\beta)=1$ by (\ref{14}). We now consider this sign factor in more detail.
Setting $\mu\!=\!0$ in (\ref{15}) we find 
$det D_A(0)=\epsilon(A)\,\Big\lb\prod_{Re(\lambda)=0}2\,cos(\beta\,Im(\lambda)/2)\Big\rb
\Big\lb\prod_{Re(\lambda)>0}|(1+e^{-\beta\lambda})|^2\,e^{\beta\,Re(\lambda)}\Big\rb$
which shows that $\epsilon(A)$ determines the sign of $det D_A(0)$. 
(Note that the condition $Im(\lambda)\in\lb-\pi/\beta,\pi/\beta\lb$ implies 
$cos(\beta\,Im(\lambda)/2)\ge0$. If we replace $\lambda\to\lambda+i2\pi n/\beta$ then for
$Re(\lambda)\!=\!0$ the change in sign of $\e(A)$ is compensated by a change in sign of 
$cos(\beta\,Im(\lambda)/2)$, while for $Re(\lambda)>0$ the replacement leaves $\e(A)$
unchanged since $\lambda$ enters in $\e(A)$ through the factors
$e^{-i\beta\,Im(\lambda)/2}\,e^{-i\beta\,Im(-\lambda^*)/2}=e^{-i\beta\,Im(\lambda)}$.)
Formally, the sign of $det D_A(0)$ is given by the sign of the product of the real 
eigenvalues of $D_A(0)\,$; this follows from the formal relation
$det(D_A(0)-z^*)=det(D_A(0)^*-z)^*=det(\gamma_5D_A(0)\gamma_5-z)^*=det(D_A(0)-z)^*$
which shows that the non-real eigenvalues come in complex conjugate pairs. In particular,
if the real eigenvalues are $\ge0$ then $det D_A(0)\ge0$. In the present regularised
(lattice space) setting the spacial Wilson term is a positive operator; consequently,
assuming $m\ge0$, the real parts of the eigenvalues of $D_A(0)$ are positive; hence one would
expect $det D_A(0)\ge0$ and therefore $\e(A)=1$ for all gauge fields $A$.
We now show rigorously that this is indeed the case. It is a nontrivial problem to show
this directly from the definition (\ref{17}); however, it can be inferred indirectly as follows.
Modulo some (positive) inconsequential factors, the dimensionally reduced expression 
(\ref{10}) was found in Eq.(20) of \cite{simplified} to arise as the continuous time limit
$\lim_{a\to0}\ a^{NN_{\beta}}\,det D_{\alpha}^{(1)}$ where $D_{\alpha}^{(1)}$ (with
$\alpha=\mu+i\pi/\beta$) is a lattice time---lattice space version of $D(\mu)$. Now, since
$D_{\alpha}^{(1)}$ is a finite-dimensional matrix and satisfies 
$(D_{\alpha}^{(1)})^*=\gamma_5\,D_{\alpha}^{(1)}\,\gamma_5$ for $\mu\!=\!0$ 
(i.e. $\alpha\!=\!i\pi/\beta$), the preceding reasoning rigorously implies 
$det D_{\alpha}^{(1)}\ge0$. Its $a\to0$ limit must therefore also be positive, hence
$det D_A(0)\ge0$ and therefore $\e(A)=1$ as claimed. (We remark though that the sign can
be negative when $m<0$, since in this case negative real eigenvalues are possible.
This situation occurs in Lattice QCD with Wilson fermions: to study the chiral limit
$m$ is tuned to negative values and there are so-called ``exceptional configurations''
for which the Dirac operator then has negative real eigenvalues.)

The large $\beta$ expressions (\ref{16a})--(\ref{16b}) have a 
remarkably simple form. Besides the natural factor $e^{-\beta E_{sea}}$,
and the sign factor $\e(A)$, which is positive for all $A$
when $m\ge0$ but which can be negative in certain gauge backgrounds when $m<0$,
the $\mu=0$ expression (\ref{16b}) consists only of a simple factor associated with the 
purely imaginary $\lambda$'s, while in the $\mu\ne0$ case (\ref{16a}) 
there is a further simple factor associated with the $\lambda$'s for which
$|Re(\lambda)|<|\mu|$. In particular, we see the following: {\em The fermion determinant
at large $\beta$ is independent of $\mu$ for $|\mu|<\mu_o(A)$ where}
\be 
\mu_o(A):=min\,\{|Re(\lambda)|\}\,.
\label{18}
\ee
This property was also found in \cite{Cohen} from an expression for the ratio
$det D_A(\mu)/det D_A(0)$ derived via the Matsubara summation method. By considering 
(a suitable integral over) the pion propagator, $min\{\mu_o(A)\}$ can be identified 
with $m_{\pi}/2$ \cite{Cohen} (where the minimum is taken over the statistically significant
gauge fields; similar considerations are standard in the Lattice QCD 
setting; see, e.g., \cite{Gibbs}.) It follows that the zero temperature QCD partition
function is independent of $\mu$ for $\mu<m_{\pi}/2$, as discussed at the beginning of this
paper.
The results (\ref{16a})--(\ref{16b}) reproduce the determinant ratio expression
Eq.~(6) of \cite{Cohen} (where it was implicitly assumed that $Re(\lambda)\ne0$ for 
all $\lambda$), and extend that result to the fermion determinant itself, revealing the 
presence of gauge field-dependent factors $e^{-\beta E_{sea}(A)}\,\epsilon(A)$ which cancel 
out in the determinant ratio. Having obtained the large $\beta$ expression for the fermion
determinant itself we are now in a position to investigate the subtle cancellation
in the functional integral over gauge fields in (\ref{2}), 
which, as discussed earlier, is required to explain the 
$\mu$-independence the zero-temperature QCD partition function 
for $m_{\pi}/2<\mu<m_N/3$. We hope to gain more insight into 
this in future work via the program outlined earlier.

%In summary, we have have derived a ``dimensionally reduced'' expression (\ref{10}) 
%for the QCD fermion determinant at finite temperature and chemical potential which 
%reduces to a remarkably simple form (\ref{16a})--(\ref{16b}) in the low temperature
%limit. A program for using this to gain insight into 

Besides the aforementioned program for gaining insight into the QCD phase transition at 
zero temperature and finite density there are other places where the
results obtained here may be useful. For example, it would be interesting to see,
if possible, how the high density effective theory (HDET) of Hong and Hsu
\cite{HongHsu}, where the effective fermion determinant is positive,
emerges from the dimensionally reduced determinant expression (\ref{10}).
This may shed light on possible limitations of that approach.
We conclude by mentioning another setting where the method introduced 
in this paper for obtaining dimensionally reduced determinant
expressions via a (partial) zeta-regularisation, using the formula Theorem 1 of \cite{BFK},
may also have interesting applications.
Following the Randall--Sundrum proposal 
\cite{Randall} there has been much interest in gauge theory models where 
the spacetime includes a fifth dimension interval, with the geometry of the interval
being warped so that the spacetime is a slice of $AdS_5$.
In particular, a model has recently been proposed in this framework where the fermion masses
arise not from the condensation of a Higgs field (the standard mechanism) but as a 
consequence of the boundary conditions on the fifth dimension interval \cite{Csaba}. 
It would be very interesting to derive a dimensionally reduced expression for the fermion
determinant in this model -- this would reveal a non-local 4-dimensional fermion theory
to which the 5-dimensional model is equivalent. (This proposal is reminiscent of the derivation 
of the overlap Dirac operator in lattice gauge theory \cite{Neu(PLB)} from a dimensional 
reduction of the fermion determinant of a 5-dimensional domain wall model \cite{Neu(PRD)}.
Since the overlap Dirac operator has proved to be very interesting and useful both
conceptually and for practical purposes, one may expect the same for the non-local
Dirac operator in 4 dimensions obtained from a dimensionally reduced expression for the
fermion determinant of the 5-dimensional model of \cite{Csaba}.) The boundary conditions
there may be too complicated for a simple application of the determinant formula of
\cite{BFK} such as the one in the present paper though; the result
of \cite{BFK} may first need to be extended to more general boundary conditions.
(We remark that more general boundary conditions have been considered
in \cite{Forman}, although the results there are mostly for determinant ratios.)
The effect of the warped geometry of the $AdS_5$ slice on the fermion determinant will also 
need to be dealt with, although this appears not so difficult. 
We also remark that in light of the AdS/CFT correspondence \cite{AdS} the dimensionally
reduced expression for the fermion determinant in this model can be expected to contain
CFT structure and it would be interesting to uncover this.

I thank Pierre van Baal for discussions related to this work and feedback on the manuscript,
Tom Cohen for feedback, Philippe de Forcrand for feedback and pointing out 
\cite{HongHsu}, and Xiang-Qian Luo for pointing out \cite{Luo}. I also thank the referee
for a useful comment.
The author is supported by the European Commission, contract HPMF-CT-2002-01716.

\end{document}